\documentclass[letterpaper,  %a4paper
              ]{jacow}
\makeatletter%
	\ifboolexpr{bool{xetex}}
	 {\renewcommand{\Gin@extensions}{.pdf,%
	                    .png,.jpg,.bmp,.pict,.tif,.psd,.mac,.sga,.tga,.gif,%
	                    .eps,.ps,%
	                    }}{}
\makeatother

\ifboolexpr{bool{xetex} or bool{luatex}} % test for XeTeX/LuaTeX
 {}                                      % input encoding is utf8 by default
 {\usepackage[utf8]{inputenc}}           % switch to utf8

\usepackage[USenglish]{babel}			 

\usepackage[final]{pdfpages}
\usepackage{multirow}
\usepackage{ragged2e}

\usepackage{siunitx}

\usepackage{float}

\begin{document}

\title{Stochastic Cooling Enhanced\\ Steady-State Microbunching}

\author{ Xiujie Deng\thanks{dengxiujie@mail.tsinghua.edu.cn\\
	Presented in 15th International Workshop on Beam Cooling and Related Topics, COOL'25, Stony Brook, New York, USA. Oct. 26-31, 2025.
}, Institute for Advanced Study,
	 Tsinghua University, Beijing, China 
	}
	
\maketitle

\begin{abstract}
   In this paper, we propose to combine two promising research topics in accelerator physics, i.e., optical stochastic cooling (OSC) and steady-state microbunching (SSMB). Our study shows that such an OSC-SSMB storage ring with a circumference of 50~m and beam energy of several hundred MeVs using present technology can deliver kilowatt radiation at 100 nm wavelength. A more ambitious application of OSC in an SSMB ring can push the radiation wavelength to an even shorter wavelength, such as EUV and soft X-ray. Such a powerful compact light source could benefit fundamental science research and industry applications. 
\end{abstract}

\section{Introduction}

Steady-state microbunching (SSMB)~\cite{Ratner2010,DengSpringer2024,Tang2026Review} scales the bunching mechanism in a storage ring from the conventional microwave or radio-frequency region to optical wavelengths to generate ultrashort electron bunches on a turn-by-turn basis for high-power short-wavelength coherent radiation generation, and its proof-of-principle experiment has been successfully conducted recently at the MLS storage ring~\cite{Deng2021,Kruschinski2024}. Optical stochastic cooling (OSC)~\cite{OSC1993,OSC1994} is a scaling of the conventional stochastic cooling scenario from the microwave to optical frequency range to speed up the damping of particle beam emittance. Its mechanism has also been demonstrated recently in the IOTA storage ring~\cite{OSCNature2022,OSC2021}. One interesting idea is then to combine them, which hopefully can relax the technical requirements and enhance the capabilities of an SSMB radiation source. 

\begin{figure}[tb]
	\centering
	\includegraphics[width=1\linewidth]{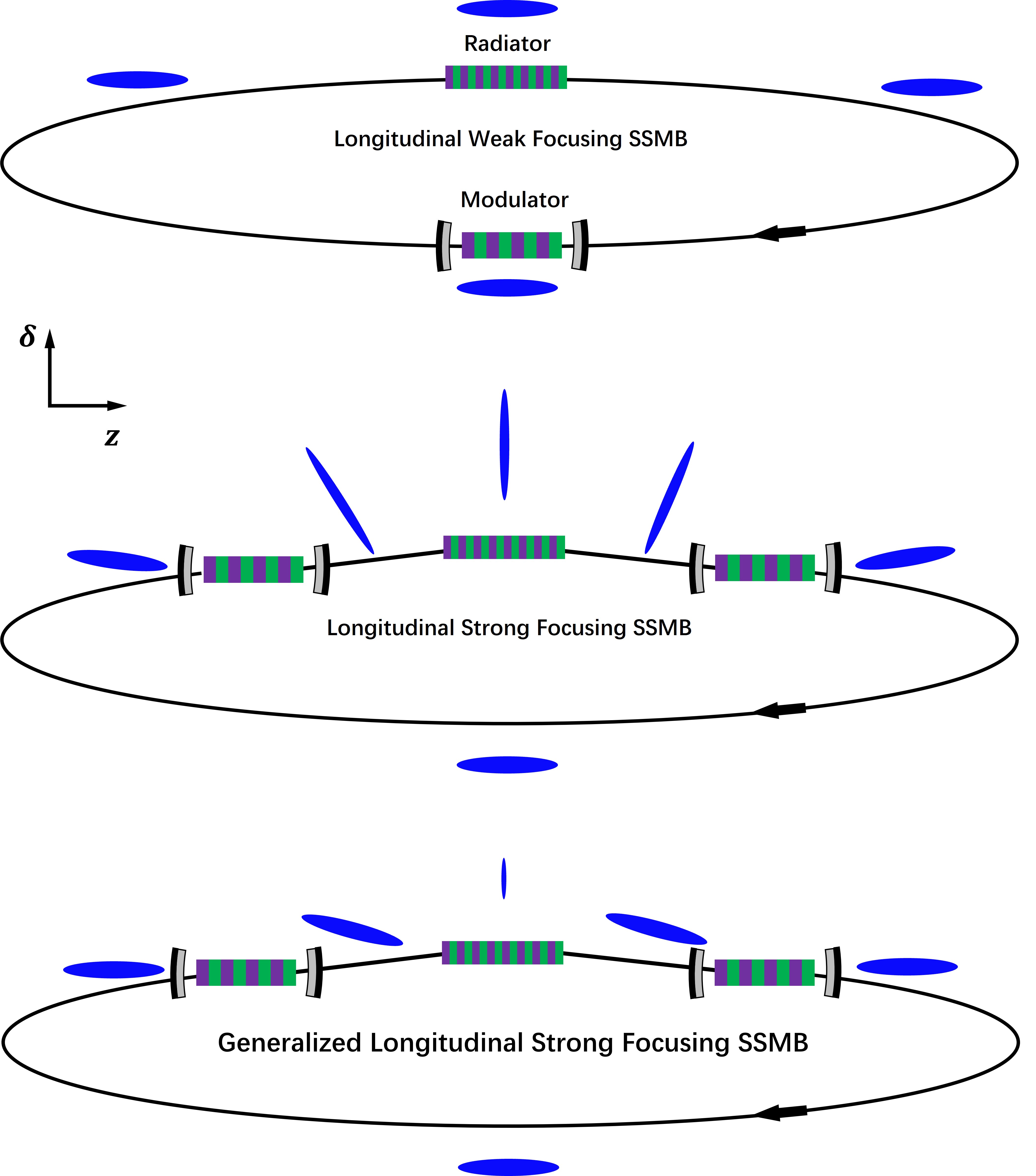}
	\caption{Schematic layouts of a longitudinal weak focusing (top), longitudinal strong focusing (center) and generalized longitudinal strong focusing (bottom) SSMB storage ring.}
	\label{fig:threessmb}
\end{figure}

\section{SSMB Scenarios}

SSMB is a general concept and there are several specific scenarios for its realization. Here we group these scenarios into two categories, i.e., globally microbunching schemes and locally microbunching schemes.

\subsection{Globally Microbunching / Longitudinal Focusing}

For globally microbunching or longitudinal focusing schemes, it means the electron beam is microbunched all around the ring. Generally, these SSMB schemes require the storage ring to work in a quasi-isochronous or low-alpha mode. The laser modulators are used in a way similar to that of RF cavities in a conventional storage ring, i.e., to longitudinal focus the electron beam to make it become microbunched. The microbunches are thus separated with a distance of the modulation laser wavelength.  Note that due to the impact of local phase slippage factor and transverse-longitudinal coupling~\cite{DengSpringer2024}, the microbunch length can vary significantly around the ring. The microbunching here therefore more accurately refers to microbunching in phase space. Depending on the strength and mechanism of longitudinal focusing, we have developed three such SSMB scenarios in the past years, namely longitudinal weak focusing (LWF), longitudinal strong focusing (LSF) and generalized longitudinal strong focusing (GLSF)~\cite{LIGLSF2023,Deng2026SSMBNST}, as schematically shown in Fig.~\ref{fig:threessmb}. For a comprehensive analysis of their beam physics, the readers can refer to Ref.~\cite{Deng2026SSMBNST}. 

For globally microbunching schemes, the key is to realize a small equilibrium beam emittance. At least one of the three eigen emittances should be ultra-low to realize ultra-short electron bunch with a mild requirement on the modulation laser power. For LWF and LSF SSMB, it is the longitudinal emittance, while for GLSF SSMB it is the vertical emittance.

\begin{figure}[tb]
	\centering
	\includegraphics[width=1\linewidth]{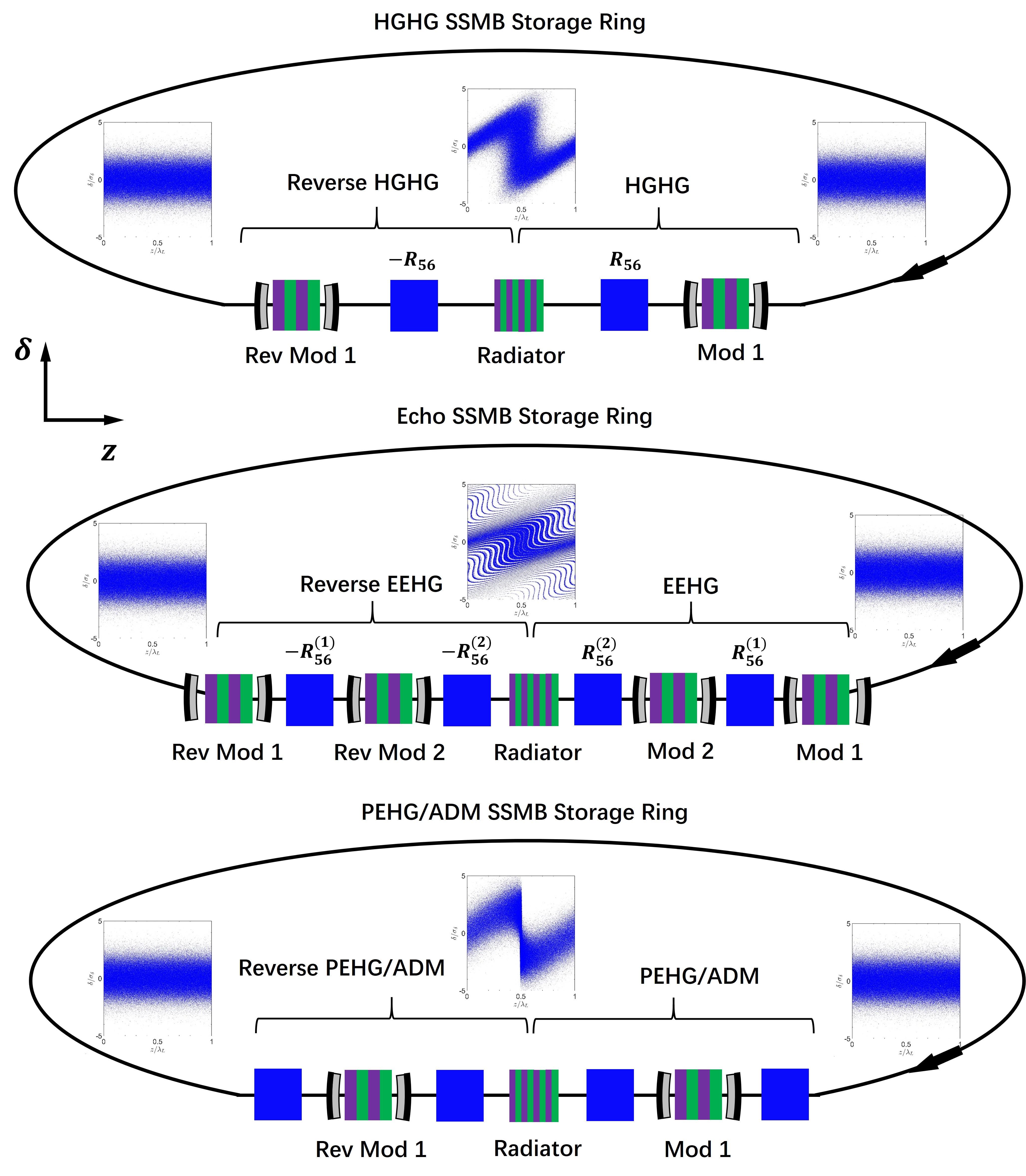}
	\caption{Schematic layouts of a reversible HGHG (top), reversible EEHG (center) and reversible PEHG/ADM (bottom) SSMB storage ring.}
	\label{fig:threermb}
\end{figure}

\subsection{Locally  Microbunching / Reversible Modulation }

For locally microbunching schemes, it means microbunching only appears in a limited section in a storage ring. Outside that limited region, the electron beam is just a conventional bunch, which means it can be an RF bunch, or even a coasting beam. A representative method to realize locally microbucnhing is using a downstream reverse modulation to cancel the modulation imprinted in the upstream laser modulator, and microbunching only appears at the radiator in between~\cite{Ratner2011Reversible,Deng2026EchoSSMB}. By invoking such a reversible modulation scheme, the storage ring does not need to be quasi-isochronous. Depending on the laser-induced microbunching technique, for example HGHG~\cite{Yu1991HGHG}, EEHG~\cite{Stupakov2009EEHG}, PEHG/ADM~\cite{Deng2013PEHG,Feng2017ADM}, we can have HGHG SSMB, Echo SSMB, PEHG/ADM SSMB as shown in Fig.~\ref{fig:threermb}.  For a comprehensive analysis of beam physics of these reversible microbunching schemes, especially that of the Echo SSMB scenario, the readers can refer to Ref.~\cite{Deng2026EchoSSMB}.

The key of reversible microbunching is to realize perfect modulation cancellation such that the equilibrium beam parameters can preserve and the microbunching process can repeat turn-by-turn. However, in reality, due to various physical effects, the modulation cancellation cannot be perfect. One critical issue is the longitudinal coordinate deviation of a particle $\Delta z$ from its ideal location between the upstream modulator and downstream reverse modulator, 
\begin{equation}
\Delta\delta=\frac{h}{k_{L}}\{\sin[k_{L}(z+\Delta z)]-\sin(k_{L}z)\},
\end{equation}
with $k_{L}$ the laser wavelength and $h$ the energy chirp strength around zero-crossing phase.
The sources of $\Delta z$ include effects like quantum excitation, intrabeam scattering (IBS), coherent  radiation, linear optics mismatch and lattice nonlienarities~\cite{Deng2026EchoSSMB}. 
A non-perfect cancellation will result in a growth of the energy spread, and when $|\Delta z|\ll\lambda_{L}$, we have
$
\Delta\sigma_{\delta}^{2}={(h\sigma_{\Delta z})^{2}}/{2}.
$
And if the modulators are placed at dispersive locations, like that in PEHG/ADM SSMB, it will also lead to a growth of transverse emittance.

For HGHG and Echo SSMB, the key parameter is the energy spread. For PEHG/ADM SSMB, it is the vertical emittance. A growth of these parameters means the required modulation laser power will be even higher to realize a given coherent radiation wavelength.  Our analysis has revealed a universal criterion for the tolerance of rms value of $\Delta z$~\cite{Deng2026EchoSSMB}
\begin{equation}
\sigma_{\Delta z}\lesssim\frac{\lambda_{R}}{\pi}\sqrt{{2}/{N_{z,\text{damp}}}},
\end{equation}
where $\lambda_{R}$ is the radiation wavelength, and $N_{z,\text{damp}}$ is the longitudinal radiation damping in unit of revolution numbers.

\subsection{Motivation of a Faster Damping}

As can be seen from the introduction above, a faster damping in an SSMB storage ring can help to obtain a smaller equilibrium beam emittance, energy spread, and increase the tolerance of non-perfect modulation cancellation. This can mitigate the technical challenges and enhance the potential of an SSMB source. While damping wiggler seems to be a straightforward choice to enhance damping, an OSC section has the advantage of a more compact setup and a higher efficiency of wall-plug electricity to user desired radiation.

\iffalse

The laser modulation system of SSMB usually consists of an optical enhancement cavity (OEC) together with an undulator magnet. The upper limit of average laser power stored in the OEC is now about 1~MW~\cite{Lu2025OEC}. To ensure large enough energy modulation depth, the beam energy cannot be too high. The typical beam energy of an SSMB storage ring is a couple of 100 MeV. To get high-average-power radiation output, we also hope for a high average beam current. These features combined lead to the fact that collective effects like intrabeam scattering (IBS) is an important issue.

To fight against collective effects like IBS and increase tolerance of non-perfect modulation cancellation, we can speed up damping. One natural idea is to use damping wiggler, which is a effective but costive solution.

\fi

\section{OSC Basics}

%Some basics of OSC can be found in Ref.~\cite{Deng2023OSC}.

After a brief introduction of SSMB, here we present some basics of OSC to make our paper more self-contained.

\subsection{Cooling Mechanism and Damping Rate}
The cooling mechanism of an OSC section is to use each electron’s radiation generated in the pick-up
undulator to correct its own momentum deviation in the kicker undulator. The
kickes of nearby electrons’ radiations in the radiation slippage length is a heating effect.
Assume that the corrective energy kick due to each particle's own radiation is given by 
%change of a particle’s relative energyinduced in the kicker undulator due to its own radiation at the pick-up undulator is
\begin{equation}\label{eq:OSCkick}
\Delta\delta_{\text{O}}=-A\sin(k_{R\text{O}}\Delta z_{\text{O}}).
\end{equation} 
And if $\Delta z_{\text{O}}=R_{56}\delta$ with $R_{56}$ the momentum compaction between the pick-up and kicker undulator, then the longitudinal damping rate is
\begin{equation}
\alpha_{L\text{O}}={Ak_{R\text{O}}R_{56}}/{2}.
\end{equation}
For the analysis of damping rates in a general coupled lattice, the readers can refer to Ref.~\cite{Deng2023OSC}. Considering the sinusoidal waveform of the kick, the damping rate is actually amplitude-dependent. And to ensure damping for majority of the particle beam, we need a sufficient cooling range~\cite{OSC2021}. 

\subsection{Bandwidth Limit}

The bandwidth $W$ of an OSC section depends on the ability of identifying each electron, which is determined by the OSC undulator radiation slippage length $N_{u\text{O}}\lambda_{R\text{O}}$. So 
$
W={f_{R\text{O}}}/{N_{u\text{O}}},
$
with $f_{R\text{O}}$ and $N_{u\text{O}}$ the OSC undulator radiation frequency and undulator period number, respectively.
%The shorter the radiation slippage length $N_{u}\lambda_{R}$, thus the better ability to identify each electron, the larger the bandwidth. Optical wavelength is much shorter than that of the microwave, thus can realize a much larger bandwidth. 
Assume perfect mixing and appropriate amplifier applied, the theoretical maximal damping rate is $\frac{1}{\tau}=\frac{W}{N}$, with $N$ the total number of particles in the ring. For a coasting beam, we have the optimal damping time in unit of revolution
\begin{equation}
N_{z,\text{damp},\text{opt}} = \text{No. of $e$ in the OSC radiation slippage}.
\end{equation}
To get a fast damping in OSC, a short radiation slippage length is required. In this sense, a shorter OSC undulator radiation wavelength is desired, for example EUV~\cite{Zholents2021OSC}. 

%Put in some numbers to get a feeling, if $\lambda_{R}=200$ nm, $N_{u}=5$, $I_{P}=1$~A, then the number of electrons in the slippage length is $N_{s}=\frac{I_{P}N_{u}\lambda_{R}}{ec}=2.08\times10^{4}$.

\subsection{Mixing Condition}

A central problem in stochastic cooling is the mixing. Loosely, a good mixing means the particles in the OSC radiation slippage length update turn-by-turn. More accurately, it means the overlap of particles' Schottky bands in the feedback system bandwidth. Since reversible SSMB scenarios do not require a small momentum compaction for the ring, the mixing condition can be straightforwardly satisfied in these schemes, as long as we have
\begin{equation}
|\sigma_{\delta}\eta C_{0}|\gg N_{u\text{O}}\lambda_{R\text{O}},
\end{equation}
with $\sigma_{\delta}$ the energy spread, $\eta$ and $C_{0}$ the phase slippage factor and circumference of the ring.
In this paper we only discuss the application of OSC in a reversible SSMB ring.

The mixing condition in globally microbunching schemes is more subtle. And the combination of these schemes with OSC will be reported elsewhere in the future.

\subsection{Radiation Kick Strength}

The OSC radiation kick strength $A$ in Eq.~(\ref{eq:OSCkick}) is determined by the details of pick-up, kicker, optical system and amplifier if there is one.  
Assume identical pick-up and kicker planar undulator, and the optical system is refractive. Assume a perfect linear amplifier. The radiation kick strength is~\cite{OSC2021}
\begin{equation}\label{eq:kick}
\begin{aligned}
A
%&=\frac{1}{4\pi\varepsilon_{0}}\frac{\frac{\pi}{3}e^{2}k_{R}N_{u}}{\gamma m_{e}c^2}F_{T}(K,\gamma\theta_{m})\\
&=\frac{1}{4\pi\varepsilon_{0}}\frac{\left(e\gamma K k_{u}\right)^{2}}{3\gamma m_{e}c^{2}}2L_{u}[JJ]F_{h}(K,\gamma\theta_{m})\sqrt{G},
\end{aligned}
\end{equation}
with $\theta_{m}$ the angular acceptance of the focusing lens, $G$ the radiation power amplification factor and for $0\leq K\leq4$ we have
$
F_{h}(K,\infty)\approx {1}/{\left(1+1.13K^2+0.04K^3+0.37K^4\right)} .
$
%Note that for radiation wavelength shorter than 200 nm, there may be no appropriate amplifier available.	

\begin{figure}[b]
	\centering
	\includegraphics[width=1\linewidth]{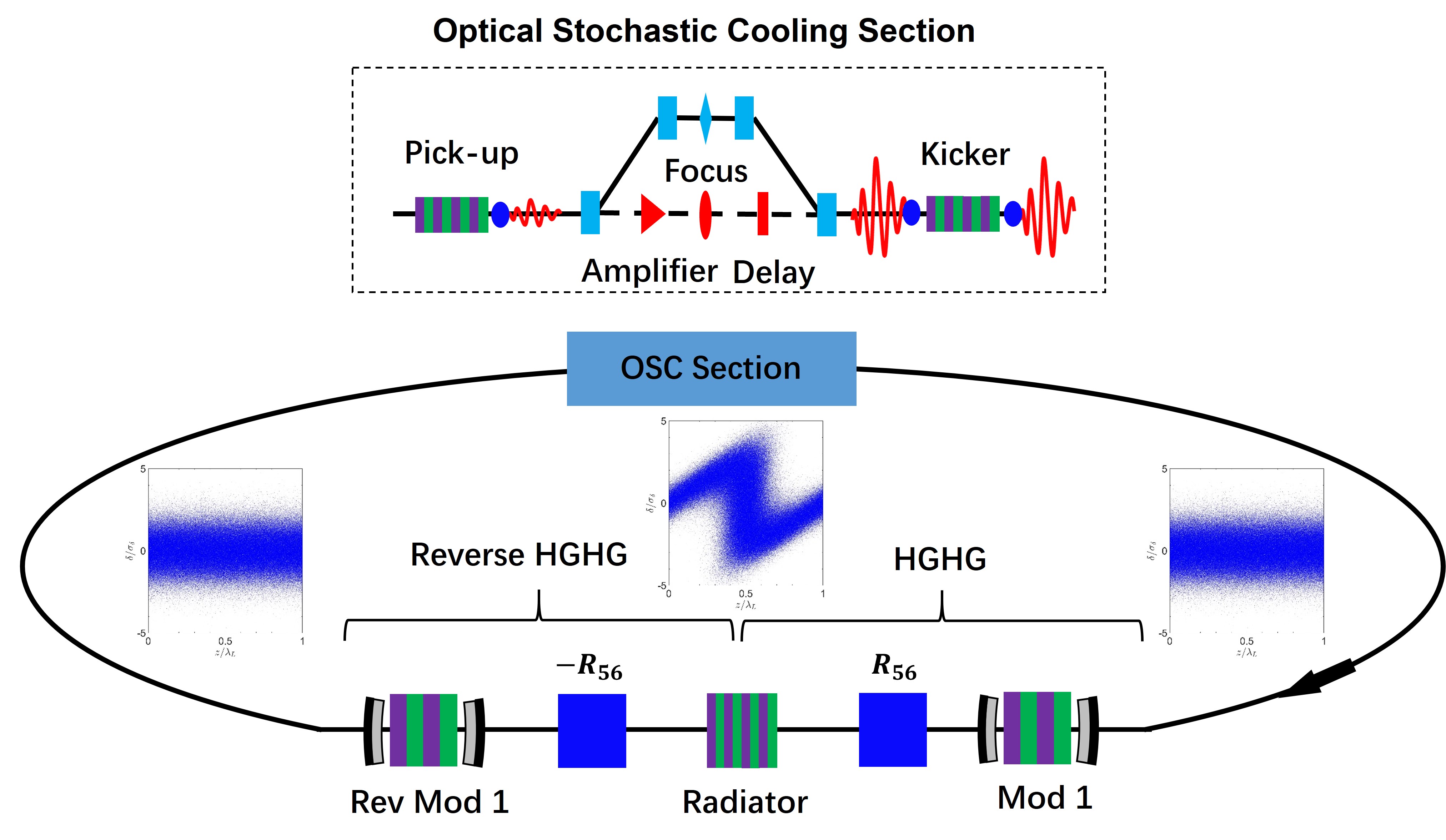}
	\caption{Schematic layout of applying OSC in a reversible HGHG SSMB storage ring.}
	\label{fig:OSCHGHGSSMB}
\end{figure}

\subsection{Equilibrium Energy Spread}

%If we consider the combined affects of radiation damping, quantum excitation and other diffusion, 
%\begin{equation}
%\begin{aligned}
%\frac{d\sigma_{\delta}^2}{dn}&=-2\alpha_{L\text{RD}}\sigma_{\delta}^2+2\alpha_{L\text{RD}}\sigma_{\delta0}^2+\Delta\sigma_{\delta}^{2}/2,
%\end{aligned}
%\end{equation}
%The equilibrium energy spread is given by
%$
%\sigma_{\delta}=\sigma_{\delta0}\sqrt{1+\frac{\Delta\sigma_{\delta}^{2}}{4\alpha_{L\text{RD}}}},
%$
%with $\sigma_{\delta0}=\sqrt{\frac{C_{q}}{J_{s}}\frac{\gamma^{2}}{\rho}}$ the natural energy spread given by radiation damping and quantum excitation.
%If we want OSC to be effective, then we need the equilibrium energy spread given by OSC alone be smaller than that determined without the OSC.

If we consider the combined affects of OSC, radiation damping, quantum excitation and other diffusion, the equilibrium energy spread is given by,
\begin{equation}
\sigma_{\delta\text{O}}=\frac{1}{2}\sqrt{{\left(\Delta\sigma_{\delta\text{O}}^{2}+\Delta\sigma_{\delta\text{QE}}^{2}+\Delta\sigma_{\delta\text{other}}^{2}\right)}/{\left(\alpha_{L\text{O}}+\alpha_{L\text{RD}}\right)}}.
\end{equation}
%\begin{equation}
%\begin{aligned}
%\frac{d\sigma_{\delta}^2}{dn}&=-2\left(\alpha_{L\text{O}}+\alpha_{L\text{RD}}\right)\sigma_{\delta}^2\\
%&+\Delta\sigma_{\delta,\text{O}}^{2}/2+\Delta\sigma_{\delta,\text{QE}}^{2}/2 +\Delta\sigma_{\delta,\text{other}}^{2}/2,
%\end{aligned}
%\end{equation}
where $\Delta\sigma_{\delta\text{O}}^{2}=\frac{I_{P}}{ec}\frac{N_{u\text{O}}\lambda_{R\text{O}}}{3}A^{2}$ is the heating effect of nearby particles' radiation kick in the OSC,  $\Delta\sigma_{\delta\text{QE}}^{2}=4\alpha_{L\text{RD}}\sigma_{\delta0}^2$ is the natural quantum excitation with $\sigma_{\delta0}$ the natural energy spread and $\alpha_{L\text{RD}}$ the longitudinal radiation damping rate, $\Delta\sigma_{\delta\text{other}}^{2}$ is from the other energy spread growth effects, for example that due to the non-perfect cancellation in a reversible SSMB ring. With other parameters fixed, there is an optimal $A$ to minimize the equilibrium energy spread. 

%If we want OSC to be effective, then we need this equilibrium energy spread smaller than that determined without the OSC section.

\section{Combine OSC with Reversible SSMB}\label{sec:longitudinal}

\subsection{Schematic Setup and Parameters List}

Now we use the application of OSC in a reversible HGHG SSMB ring to increase the longitudinal damping rate as an example to demonstrate the benefit of implementing OSC in an SSMB storage ring. A schematic layout of such an OSC-HGHG-SSMB storage ring is shown in  Fig.~\ref{fig:OSCHGHGSSMB}, and an example parameters list is given in Tab.~\ref{tab:LWFSSMBOSC}. In this example, the desired radiation wavelength is assumed to be the 8-th harmonic of the modulation laser, i.e., $\lambda_{R}=100$~nm with $\lambda_{L}=800$ nm. The envisioned ring consists of two arcs and two straight sections. The circumeference of such a ring is about 50~m.  The two straights are about $2\times15$ m, and the two arcs are  about $2\times10$ m.  The OSC section is implemented in one straight section, while the other is used for HGHG and reverse HGHG. The average modulation laser power is set to be 1 MW, which is close to the state-of-art value reachable in an optical enhancement cavity~\cite{Lu2025OEC}.  All the parameters in Tab.~\ref{tab:LWFSSMBOSC} should be feasible from a practical viewpoint. 

%In the parameters choice we have made the cooling range $\frac{\mu_{01}}{k_{R\text{O}}R_{56}\sigma_{\delta\text{O}}}=3$, where $\mu_{01}=2.405$ is the first root of the zero-th order Bessel function of the first kind. 

\begin{table}[tb]
	\caption{\label{tab:LWFSSMBOSC}
		An example parameters set of an OSC-HGHG-SSMB ring for high-power 100 nm radiation generation.}
	\centering
	%\begin{ruledtabular}
	\begin{tabular}{lll}  
		\hline
		Para. & Value & Description \\
		%\colrule
		\hline 
		$E_{0}$ & $300$ MeV & Beam energy \\
		$C_{0}$ & 50 m & Circumeference \\
		$\eta$ & $1\times10^{-3}$ & Phase slippage factor \\	
		$I_{P}$ & 1 A & Peak current\\
		%$f$ & 50\% & Beam filling factor\\
		$I_{A}$ & 0.5 A & Average current\\					
		%$\rho_{\text{ring}}$ & 0.8 m & Bending radius of dipoles \\
		$B_{\text{ring}}$ & 1.25 T & Bending field of dipoles \\
		$U_{0}$ & 896 eV & Radiation loss in dipoles \\ 
		$\sigma_{\delta0}$ &  $2.9\times10^{-4}$ & Natural energy spread\\
		$\tau_{\delta\text{RD}}$ & 55.8 ms & Longitudinal R.D. time\\
		%$\sigma_{\delta\text{NoOSC}}$ &  $>4.8\times10^{-4}$ &  Energy spread with diffusion\\
		%$\sigma_{z{\text{RD}}}$ &  68.5 nm & Natural bunch length\\
		%$\epsilon_{z\text{RD}}$ & 23.5 pm & Natural longitudinal emittance\\
		$\frac{\alpha_{L\text{O}}}{\alpha_{L\text{RD}}}$ & 10 & Ratio of two damping\\
		$\sigma_{\delta{\text{O}}}$ &  $1.9\times10^{-4}$ &  Energy spread with OSC\\					
		$\tau_{\delta\text{O}}$ & 5.1 ms & Damping time with OSC\\	
		%$N_{z,\text{damping}}$ & $3\times10^{4}$ & Damping time in revolution number\\				
		%					$\sigma_{z{\text{O}}}$ &  20 nm & Bunch length at radiator with OSC\\
		%					%$\epsilon_{z\text{O}}$ & 2 pm & Emittance with OSC\\
		%					$\sigma_{z,\text{lim}}$ & 7.6 nm & Bunch length limit with OSC\\
		$\tau_{\delta,\text{IBS}}$ & $\gtrsim100$ ms & IBS diffusion time\\
		%$\rho_{\text{ring}}$ & 0.6 m & Bending radius in the ring \\
		%$\theta$ & $\frac{\pi}{10}$ & Bending angle of each dipole \\	
		%$\sigma_{z,\text{lim}}$ & 26 nm & Theoretical bunch length limit \\
		%$\epsilon_{x}$ & 3 nm &Horizontal emittance\\
		%$\epsilon_{z,\text{lim}}$ & 32 pm & Theoretical minimum longitudinal emittance \\
		%$\lambda_{L}$ & 1064 nm & Modulation laser wavelength \\		
		\hline
		$\lambda_{L}$ & 800 nm & Laser wavelength \\
		%$V_{L}$ &  360 kV & Laser induced modulation voltage \\ 
		%$A_{L}$ &  $1.2\times10^{-3}$ & Laser induced modulation  \\ 		
		$h$ &  $9425\ \text{m}^{-1}$ & Liner energy chirp strength \\ 
		$\lambda_{u}$ & 4.5 cm & Modulator undulator period\\
		$B_{0}$ & 1.13 T & Modulator peak field\\
		%$K_{u}$ & 4.74 & Undulator parameter\\
		$L_{u}$ & 3.15 m & Modulator undulator length\\
		%$R_{y}$ & ${L_{u}}/{3}$ & Rayleigh length\\
		%$P_{L,P}$ &  2 MW & Laser peak power\\ 
		$P_{L,A}$ &  1 MW & Laser average power\\ 
		\hline 					
		%Passive OSC & & \\
		$\sigma_{\Delta z}$ & 0.2 nm & rms $\Delta z$ between LMs\\
		$\Delta\sigma_{\delta\text{ins}}^{2}$ & $1.8\times10^{-12}$ &  $\Delta\sigma_{\delta}^{2}$ per pass of insertion\\
		$\Delta\sigma_{\delta\text{O}}^{2}$ & $1.9\times10^{-12}$ &  $\Delta\sigma_{\delta}^{2}$ per pass of OSC\\
		$\Delta\sigma_{\delta\text{QE}}^{2}$ & $1\times10^{-12}$ &  $\Delta\sigma_{\delta}^{2}$ quantum excitation\\
		\hline 					
		%Passive OSC & & \\
		$\lambda_{R\text{O}}$ & 266 nm & OSC radiation wavelength\\
		$\lambda_{u\text{O}}$ & 3 cm & OSC undulator period\\
		$B_{0\text{O}}$ & 1.14 T & OSC undulator field\\
		%$K_{u\text{O}}$ & 3.2 & OSC undulator parameter\\
		%$N_{u}=5$ & 5 & Number of undulator period\\
		$L_{u\text{O}}$ & 15 cm & OSC undulator length\\
		$N_{s}$ & $2.77\times10^{4}$ & No. of $e$ in a slippage length\\
		$G$   & 178 & Power amplification factor \\
		$R_{56}$ & 179 $\mu$m & $R_{56}$ between two undulators\\
		$\frac{\mu_{01}}{k_{R\text{O}}R_{56}\sigma_{\delta\text{O}}}$ & 3 & Cooling range \\
		%$\frac{\mu_{01}}{k_{R}F\sigma_{\delta\text{O}}}$ & 3 & Cooling range \\
		%		\hline
		%		Passive OSC & & \\
		%		$\lambda_{\text{O}}$ & 266 nm & OSC undulator radiation wavelength\\
		%		$\lambda_{u}$ & 3.5 cm & Modulator undulator period\\
		%		$B_{0}$ & 1.25 T & Modulator peak magnetic field\\
		%		$K_{u}$ & 4 & Undulator parameter\\
		%		$L_{u}$ & 3.5 m & Undulator length\\
		%		%$G$   & 77 & Power amplification factor \\
		%		$R_{56}$ & 339 $\mu$m & $R_{56}$ between two undulators\\
		\hline
		$\lambda_{R}$ & 100 nm & Radiation wavelength\\
		$b_{8}$ & 0.1 & Bunching factor\\
		$\epsilon_{\bot}$ & 6 nm & Transverse emittance\\
		$\sigma_{\bot}$ & $\sim100$ $\mu$m & Transverse beam size\\
		$\lambda_{u}$ & 2 cm & Radiator undulator period\\
		$B_{0}$ & 1.18 T & Radiator peak field\\
		%$K_{u}$ & 2.2 & Undulator parameter\\
		$N_{u}$ & 300 &  Undulator period number\\
		$L_{u}$ & 6 m & Undulator length\\
		$P_{P}$ &  1 kW & Peak radiation power\\ 
		$P_{A}$ &  0.5 kW & Average radiation power\\
		%$\mathcal{F}_{P}$ & $>10^{19}$ phs/s/0.1\%b.w. & Peak radiation spectral flux\\
		\hline		
	\end{tabular}
	%\end{ruledtabular}
\end{table}

% We use a four-dipole achromatic chicane to realize the desired $F$ or $R_{56}$,
%\begin{equation}
%F=R_{56}=2L\theta^{2}.
%\end{equation}
%A normal four-dipole chicane is an achromat. 
%If $R_{56}=100\ \mu$m, and $\sigma_{\delta}=3.43\times10^{-4}$, then to ensure $3\sigma_{\delta}$ damping range, we need
%\begin{equation}
%\lambda_{R}>269\ \text{nm}.
%\end{equation}

%To realize the desired longitudinal emittance, we need 
%\begin{equation}
%\frac{Ak_{R}R_{56}}{2}>2.6\times10^{-5}.
%\end{equation}
%If $\lambda_{OSC}=30$ nm, 
%then $A>2.5\times10^{-8}$, 
%and if we choose the pick-up and kicker undulator parameters 

%Give a plot of $A$ as a function of $K$, with $\lambda_{R}$ and $L_{u}$ fixed. Basically, we expect the larger the $K$, the larger the $K$.
%
%With $\lambda_{R}$ and $L_{u}$ fixed, we fins that the smaller $\lambda_{u}$ is, the larger $K$ and also the larger kick strength $A$ is. So in practice, we choose as small $\lambda_{u}$ as we can.

\subsection{Induction Linac and Barrier Bucket}

\begin{figure}[tb]
	\centering
	\includegraphics[width=1\linewidth]{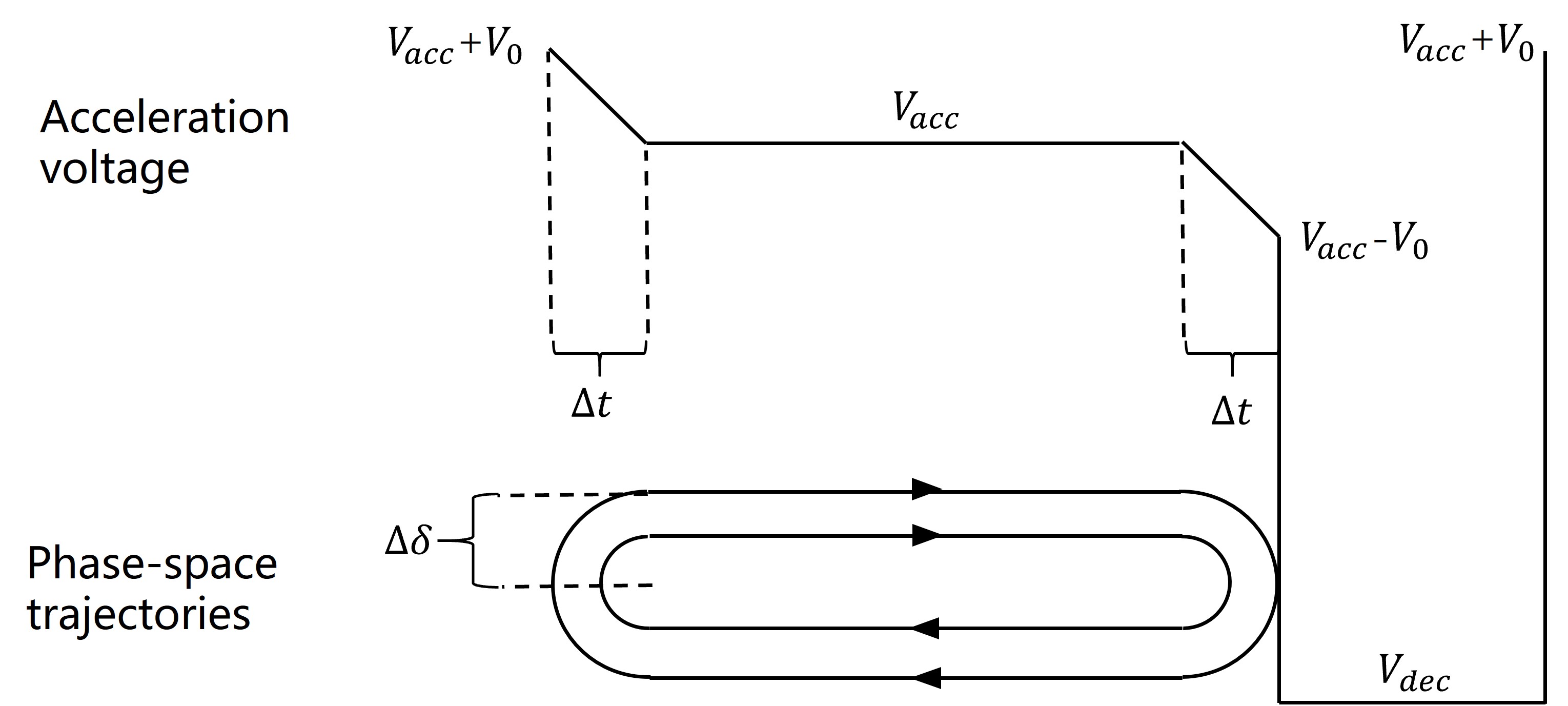}
	\caption{The schematic layout of applying an induction linac for energy compensation and barrier bucket formation.}
	\label{fig:induction}
\end{figure}

To mitigate collective effects like IBS, the peak current we applied is not too high, namely 1 A. While for a high-average-power output radiation, we hope for a high average beam current. Therefore, we may use induction linac as the energy compensation system to get a costing beam with a large filling factor for example 50\% in the ring. Assuming the repetition rate of the induction linac is the same as the particle revolution frequency in the ring, which is 6 MHz. The acceleration voltage $V_{\text{acc}}\approx2$ kV, considering that the incoherent radiation loss in dipoles and coherent radiation loss in radiator undulator are each about 1 keV per electron. To form the barrier bucket, we let the acceleration voltage have two edge slopes as shown in Fig.~\ref{fig:induction}. The barrier bucket half-height formed by these two slopes is
\begin{equation}
\Delta\delta=\sqrt{\frac{eV_{0}}{E_{0}}/\frac{\eta C_{0}}{c\Delta t}},
\end{equation}
with $V_{0}$ and $\Delta t$ the magnitude and time duration of the slope. If $E_{0}=300$ MeV, $C_{0}=50$ m, $\eta=1\times10^{-3}$, $\Delta t=20$ ns, then to realize a bucket half-height of $\Delta\delta=0.02$ which means about $100\sigma_{\delta\text{O}}$ given in the table, then we need $V_{0}=1$ kV. Such a requirement on induction linac should be feasible. We point out that it is also possible to realize a large beam filling factor by using the combination of different RF harmonic cavities to lengthen the RF bunch in a RF bucket.

% An example radiation energy spectrum and spatial distribution from the formed microbunch train with different transverse beam sizes is shown in Fig.~\ref{fig:figure2}. The average radiation flux and power will be the values presented in the figure multiplied by the beam filling factor, in our case 50\%.  As can be seen, we can realize an average power of 1.5 kW, and spectral flux $>10^{20}\ \text{phs/s/0.1\%b.w.}$ at the EUV wavelength , with an average beam current of 0.5 A. These values are  four orders of magnitudes larger than the present synchrotron sources. Such a high-flux EUV radiation source is appealing for fundamental condensed matter physics study, for example to investgate the energy gap distribution of quantum materials using the ARPES technique. More details on how the coherent undulator radiation spectrum and spatial distribution is obtained can be found in Chapter 4 of Ref.~\cite{DengSpringer2024}.

%\begin{figure}
%	%   \vspace*{-.5\baselineskip}
%	\centering
%	\includegraphics*[width=0.5\textwidth]{TUP164-THA_f2a}\\
%	\includegraphics*[width=0.32\columnwidth]{TUP164-THA_f2b}
%	\includegraphics*[width=0.32\columnwidth]{TUP164-THA_f2c}
%	\includegraphics*[width=0.32\columnwidth]{TUP164-THA_f2d}
%	\caption{Example radiation energy spectrum and spatial distribution from microbunch train, with different transverse beam sizes. The total radiation power are 16.3 kW, 5.3 kW, 2.9 kW, respectively. }
%	\label{fig:figure2}
%	%   \vspace*{-\baselineskip}
%\end{figure}

\subsection{Intrabeam Scattering}
To minimize the IBS diffusion rate, we have used a transversely round electron beam, with a large transverse emittance of $\epsilon_{\bot}=6$ nm.
The IBS diffusion of  energy spread for a transversely round beam ($\epsilon_{x}=\epsilon_{y}$) is~\cite{IBSHandbook}
$
\tau_{\delta,\text{IBS}}^{-1}\approx\frac{\Psi_{0} I_{P}r_{e}^{2} L_{C}}{8e\gamma^{3}\sigma_{\delta }^{2}\langle\sigma_{x}\rangle\epsilon_{\bot}},
$
where $\Psi_{0}$ is a constant depending on the lattice optics around the ring.
 Here for an order of magnitude estimation, we put in some numbers, $\Psi_{0}=1$, Columb Log $L_{c}=10$, average transverse beam size around the ring $\langle\sigma_{x}\rangle=\sqrt{6\ \text{nm} \times 10\ \text{m}}\approx250\ \mu$m, $\sigma_{\delta}=1.9\times10^{-4}$, $I_{P}=1$ A, we have
$
\tau_{\delta,\text{IBS}}
=169\ \text{ms},
$
which is more than one order of magnitude longer than the OSC damping time. For the transverse dimension
$
\tau_{\bot,\text{IBS}}=\frac{\epsilon_{\bot}}{\sigma_{\delta}^{2}\langle\mathcal{H}_{\bot}\rangle}\tau_{\delta,\text{IBS}}.
$
When $\langle\mathcal{H}_{\bot}\rangle\lesssim0.2$~m, it is acceptable for the transverse dimensions.

\iffalse

\subsection{Microwave Instability}
The CSR-induced microwave instability threshold bunch current with shielding for a coasting beam is~\footnote{Cai, IPAC2011-FRXAA01.} 
\begin{equation*}
\begin{aligned}
I_{b}
=\frac{3\sqrt{2}\gamma\eta \sigma_{\delta}^{2}I_{Alf}L_{b}}{\pi^{\frac{3}{2}}h},
\end{aligned} 
\end{equation*} 
where $I_{Alf}=17$ kA is the Alfven current.

Put in the numbers in our table: $E_{0}=300$ MeV,  $\eta=1\times10^{-3}$, $L_{b}=C_{0}/2=25$~m, $\sigma_{\delta0}=2\times10^{-4}$, assume gap of two parallel metal plates $h=1$ cm, then
\begin{equation*}
I_{b}=0.76\ \text{A},
\end{equation*}
which is larger than the 0.5 A average beam current we have applied.

\fi

\subsection{Coherent Undulator Radiation}

The radiator is assumed to be an undulator. The radiation power at the fundamental resonance frequency from a transversely-round electron beam in a planar undulator is~\cite{DengSpringer2024}
\begin{equation}
P_{\text{peak}}[\text{kW}]=1.183N_{u}\chi[JJ]^{2}FF_{\bot}(S)|b_{z}|^{2}I_{P}^{2}[\text{A}],
\end{equation}
%\begin{equation}
%P_{\text{rad}}=\frac{\pi}{\epsilon_{0}c}N_{u}\chi[JJ]^{2}FF_{\bot}(S)|b_{z}|^{2}I_{A}^{2}, 
%\end{equation} 
where $[JJ]^{2}=\left[J_{0}\left(\chi\right)-J_{1}\left(\chi\right)\right]^{2}$, with $\chi=\frac{K^{2}}{4+2K^{2}}$ and $K$ being the undulator parameter, and the transverse form factor is
$
FF_{\bot}(S)=\frac{2}{\pi}\left[\tan^{-1}\left(\frac{1}{2S}\right)+S\ln\left(\frac{(2S)^{2}}{(2S)^{2}+1}\right)\right],
$
with $S=\frac{\sigma^{2}_{\bot}\frac{\omega}{c}}{L_{u}}$ is the diffraction parameter and $\sigma_{\bot}$ the rms transverse electron beam size, $b_{z}$ is the bunching factor at the radiation wavelength, and $I_{P}$ is the peak current. For the radiation power of a helical undulator, there is no $[JJ]^{2}$ factor and the power can roughly be about a factor 2 larger. The power given in the table assumes a helical undulator radiator. Similarly we can also use a helical undulator as the modulator to lower the required modulation power by a factor of 2 compared to a planar undulator as modulator.

\subsection{Remaining Issues}
Feasible in principle as it is, there are still important issues to be resolved before such an OSC-HGHG-SSMB storage ring become a reality. For example, is the required amplifier of OSC radiation available at such short wavelength? How to realize the phase locking of the two optical cavities, or more accurately locking of the downstream laser modulator to the electron beam to ensure perfect modulation cancellation all the time? Is it challenging to realize the required phase-locking of electron and its own radiation in OSC?

\section{Summary}\label{sec:summary}
%OSC cooling rate and equilibrium beam parameters calculation in a 3D genral coupled lattice. 
%
%OSC cooling rate expressed more elegently using Courant-Snyder functions.
%
%Application of OSC for high-power EUV and soft X-ray generation.
%
%Application of OSC for ultrashort radiation pulse generation.
%
%Calculation of radiation properties.

Both OSC and SSMB have great potential, here we propose to combine them for an even brighter and longer future. The application of OSC in reversible SSMB scenarios is feasible and we have presented an example of 1 kW 100~nm radiation source based on this idea, using parameters all reachable using present technology. Such a compact source (circumference $\sim50$ m) can be built in universities and institutes with a reasonable cost, and be useful for basic science research. The work on application of OSC in global microbunching SSMB scenarios is ongoing and will be reported in the future. This work is supported by the National Natural Science Foundation of China (NSFC Grant No. 12522512) and Tsinghua University Dushi Program.

%We encourage readers to refer to Ref.~\cite{OSCSSMB2024} for more technical details.

%\section{ACKNOWLEDGMENTS}

%\null  % this is a hack for correcting the wrong un-indent by package 'flushend' in versions before 2015


\begin{thebibliography}{99}   % Use for  10-99  references
%\begin{thebibliography}{9} % Use for 1-9 references
	
	%\bibitem{accelconf-ref}
	%	C. Petit-Jean-Genaz and J. Poole,
	%	``JACoW, A service to the Accelerator Community,''
	%	EPAC'04, Lucerne, July 2004, THZCH03,  p.~249,
	%	\url{http://www.JACoW.org/e04/papers/THZCH03.PDF}
	
	

	
	
	
	
	
	
	
	
	\bibitem{Ratner2010}
	Ratner, Daniel F., and Alexander W. Chao. "Steady-state microbunching in a storage ring for generating coherent radiation." Physical review letters 105.15 (2010): 154801.
	
	
	
	\bibitem{DengSpringer2024}
	Deng, Xiujie. Theoretical and experimental studies on steady-state microbunching. Springer Nature, 2024.
	
	
	\bibitem{Tang2026Review}
	Tang, C. X. et al. ``Physics of steady-state microbunching." to be published in Springer Series Synchrotron Light Sources and Free-Electron Lasers.
	
	
	\bibitem{Deng2021}
	Deng, Xiujie, et al. "Experimental demonstration of the mechanism of steady-state microbunching." Nature 590.7847 (2021): 576-579.
	
	
	
	
	\bibitem{Kruschinski2024}	
	Kruschinski, Arnold, et al. "Confirming the theoretical foundation of steady-state microbunching." Communications Physics 7.1 (2024): 160.
	
	\bibitem{OSC1993}
	Mikhailichenko, A. A., and M. S. Zolotorev. "Optical stochastic cooling." Physical Review Letters 71.25 (1993): 4146.
	
	\bibitem{OSC1994}
	Zolotorev, M. S., and A. A. Zholents. "Transit-time method of optical stochastic cooling." Physical Review E 50.4 (1994): 3087.
	
	
	\bibitem{OSCNature2022}
	Jarvis, Jonathan, et al. "Experimental demonstration of optical stochastic cooling." Nature 608.7922 (2022): 287-292.
	
	\bibitem{OSC2021}
	Lebedev, V., et al. "The design of optical stochastic cooling for IOTA." Journal of Instrumentation 16.05 (2021): T05002.
	
	
	\bibitem{LIGLSF2023}
	Li, Zizheng, et al. "Generalized longitudinal strong focusing in a steady-state microbunching storage ring." Physical Review Accelerators and Beams 26.11 (2023): 110701.
	
	\bibitem{Deng2026SSMBNST}
	Deng, Xiu-Jie, et al. "Steady-state microbunching based on transverse-longitudinal coupling." Nuclear Science and Techniques 37.1 (2026): 2.
	
	
	
	\bibitem{Ratner2011Reversible}
	Ratner, Daniel, and Alex Chao. "Reversible seeding in storage rings." Proc. of the 33th International Free-electron Laser Conference. 2011.
	
	\bibitem{Deng2026EchoSSMB}
	Deng, X. J., Pan Z. L., Zhao J. Y., Chao, A. W., Tang, C. X. "Reversible microbunching in an electron storage ring." submitted to PRAB.
	
	\bibitem{Yu1991HGHG}
	Yu, Li Hua. "Generation of intense uv radiation by subharmonically seeded single-pass free-electron lasers." Physical Review A 44.8 (1991): 5178.	
	
	\bibitem{Stupakov2009EEHG}
	Stupakov, Gennady. "Using the beam-echo effect for generation of short-wavelength radiation." Physical review letters 102.7 (2009): 074801.
	
	\bibitem{Deng2013PEHG}
	Deng, H., and C. Feng. "Using off-resonance laser modulation for beam-energy-spread cooling in generation of short-wavelength radiation." Physical Review Letters 111.8 (2013): 084801-084801.
	
	\bibitem{Feng2017ADM}
	Feng, Chao, and Zhentang Zhao. "A storage ring based free-electron laser for generating ultrashort coherent EUV and X-ray radiation." Scientific reports 7.1 (2017): 4724.
	
	
	
%		\bibitem{OSCSSMB2024}
%	Deng, Xiujie. "Application of Optical Stochastic Cooling in Future Accelerator Light Sources." arXiv preprint arXiv:2407.16098 (2024).
	
	\bibitem{Deng2023OSC}
	Deng, Xiujie. "Optical Stochastic Cooling in a General Coupled Lattice." Proceedings of the 67th ICFA Adv. Beam Dyn. Workshop Future Light Sources, Luzern, Switzerland (JACoW, Geneva, 2023), p. TU4P30.
	
	
	\bibitem{Zholents2021OSC}
	Zholents, Alexander, Luca Rebuffi, and Xianbo Shi. "Stochastic cooling of electrons and positrons with EUV light." Physical Review Accelerators and Beams 24.2 (2021): 022803.

	
	
	
%	\bibitem{Chao2022}
%	Chao, A.  “Focused laser”, unpublished note, 2022.

%   \bibitem{PartialAlpha2023}
%   Deng, X. J., et al. "Breakdown of classical bunch length and energy spread formula in a quasi-isochronous electron storage ring." Physical Review Accelerators and Beams 26.5 (2023): 054001.

	
	
	
	\bibitem{Lu2025OEC}
	Lu, Xin-Yi, et al. "Stable 710kW average-power of infrared-red light stacked in an optical enhancement cavity of finesse 45,000." Frontiers in Ultrafast Optics: Biomedical, Scientific, and Industrial Applications XXV. Vol. 13353. SPIE, 2025.
	
	\bibitem{IBSHandbook}
	Lebedev, V.  Handbook of accelerator physics and engineering, 2nd ed. (World scientific, Singapore, 2013) pp. 155–159.
	
\end{thebibliography}
\end{document}